\documentclass[useAMS,usenatbib]{mn2e}
\usepackage{epsfig,amssymb}
\newcommand{\ceoneRe}{388}
\newcommand{\cetwoRe}{263}

\title[Tidal streams in M32 analogues]{Tidal streams in newly discovered M32 analogues: evidence for the stripping scenario}
\author[ ]{A. P. Huxor\thanks{E-mail:
avon.huxor@bristol.ac.uk}, S. Phillipps, J. Price and R. Harniman \\\
University of Bristol, H. H. Wills Physics Laboratory, Tyndall Avenue, Bristol, BS8 1TL, U.K. \\
}
\begin{document}

\date{}

\pagerange{\pageref{firstpage}--\pageref{lastpage}} \pubyear{2002}

\maketitle

\label{firstpage}

\begin{abstract}
We present two newly-discovered compact elliptical (cE) galaxies, exhibiting clear evidence of tidal steams, and  found during a search of SDSS DR7 for cE candidates. The structural parameters of the cEs are derived using GALFIT,  giving effective radii, R$_{e}$, of \ceoneRe{}  and \cetwoRe{}  parsecs, and B-band mean surface brightnesses within R$_{e}$ of 19.4 and 19.2 magnitudes arcsec$^{-2}$. We have re-analysed the SDSS spectra, which indicate that they possess  young to intermediate-age stellar populations. These two cEs
provide direct evidence, a "smoking gun'',  for the process of tidal stripping  that is believed to be the origin of M32-type galaxies. Both are in small groups with a large spiral fraction, suggesting that we may be seeing the formation of such cE galaxies in dynamically young environments. The more compact of the galaxies is found in a small group not unlike the Local Group, and thus provides an additional model for understanding M32.

\end{abstract}

\begin{keywords}
galaxies:formation --- galaxies:dwarf --- galaxies: structure
\end{keywords}

\section{Introduction}

There has been some debate over the origins of the so-called ``compact ellipticals".
These belong to an apparently
 rare morphological type, exemplified by M32, which has a very small effective radius ($\sim$ few hundred parsecs) and  
high central surface brightness \citep{Faber73}. There are only a handful of classical cEs, those which have been known for some decades.
In addition to M32 - a very close satellite of M31 - the other ``classical" cEs, NGC 4486B and NGC 5846A,  
 also occur close to massive galaxies in groups or clusters. 
 They are, therefore, generally suspected to be the 
result of tidal stripping and truncation driven by interactions with their giant neighbours  \citep{Faber73,KingKiser73,Bekkietal01b, Choietal02}. However, an alternative view suggests that they are the low-luminosity end of the family of normal elliptical galaxies, in which the high surface brightness is the result of earlier star formation caused by dissipative (wet) mergers \citep{Kormendyetal09}.

Recent years have seen the discovery of a number of cEs, including two in the massive galaxy cluster Abell 1689 \citep{Mieskeetal05}, another in the halo of the cD galaxy of Abell 496 \citep{Chilingarianetal07}, a sister to the classical cE in the NGC 5846 Group \citep{ChilingarianBergond10} and earlier noted in \citet{Mahdavietal05},  two in the Antlia Cluster \citep{SmithCastellietal09}, and three cEs  in the Coma Cluster \citep{Priceetal09}.

The evidence for stripping as a mechanism to form cEs is compelling but frequently  indirect. For example, they are often found to have metallicities and velocity dispersions appropriate to more massive galaxies. \citet{Kormendyetal97} found evidence that NGC 4486B possesses a central massive black hole. It has a mass well above the relation between the black hole mass and bulge luminosity found for other galaxies, consistent with a more massive progenitor. Other evidence indicates the probable Hubble type of any progenitor. For example, surface brightness profiles obtained  from  high quality imaging (such as that available with HST) are best fit by a two component model, indicating that we are seeing the remnant of a disk  \citep{Priceetal09}. Indeed,  \citet{Graham02} suggests that M32 itself still retains a disk component. Furthermore, the gas from which the stars in M32 formed is similar to that of the Galactic disk, and unlike the rapid enrichment found in classical elliptical galaxies, consistent with the threshed spiral scenario \citep{Davidgeetal08}.

Direct evidence of stripping occurring  is less forthcoming.
Suggestions of tidal features have been reported in M32 \citep{ Choietal02}, but they are difficult to identify as M32 lies close to the line of sight to M31, and the disk of M31 makes accurate photometry difficult. More recently,  both  \citet{Chilingarianetal09} and  \citet{SmithCastellietal08} report a cE exhibiting a possible tidal feature, but these features are faint and rather uncertain, again due to the high surface brightness of their immediate environments.

A number of dwarf galaxies exhibiting strong tidal features have been found, but these lie at the low luminosity end of typical early-type galaxies  (having effective radii of $\sim$1 kpc) rather than being compact ellipticals. For example, \citet{Forbesetal03} describe  the serendipitous discovery of a dwarf (M$_{V}$= -16.0 mags, $R_{e}$ = 930 pc) being stripped by a more massive neighbour. 
Another has been found in Subaru data \citep{Sasakietal07} which also has an effective radius  of $\sim$1 kpc, 
but is somewhat more luminous, with  an M$_{V}$ of -17.7 mags.                            
Tidal debris has also been seen near massive galaxies, but with no clear evidence of any progenitor, as presumably it has been totally disrupted \citep{MartinezDelgadoetal09,WehnerGallagher05}.

The lack of any clear examples of cEs being formed by stripping is not totally unexpected,
 due to the very short timescale \citep{Bekkietal01b} expected from  the stripping scenario (less than a few Gyr). 
However, identifying a number of cEs which exhibit ongoing  tidal stripping would not only show that a formation channel via tidal stripping does occur, but also allow us to study  the process in more detail. 
Equally, examples of very isolated cEs, in which stripping is not a likely formation channel would be significant, as it would suggest that cEs can form by other means, and their location close to massive galaxies may not be a direct cause of their morphology. 

Hence  we have
undertaken a search for further cE candidates from the full SDSS DR7 spectroscopic sample, which embraces galaxy cluster, group  and field regions, with the aim of finding examples that might address these issues.

\begin{table*} 
 \centering
  \caption[Basic properties of cEs. ]{ Basic properties of cEs.
  }
  \begin{tabular}{@{}llllllllll@{}}
 
SDSS ID					&  Name & RA 				&  Dec 		& $D_{L}$		& $(g-r)_{0}$	&  $D_{A}$	& Scale			& Host			& R$_{proj}$ 	 \\
						&   	&    				&            		&   Mpc		& SDSS		&  Mpc		& pc arcsec$^{-1}$ 	&         			& kpc         \\

 \hline 
 J110404.40+451618.9 	& cE1  	& 11 04 04.40 		& +45 16 18.9  &  95.2   		&  0.81 		&  91.1 		& 442 			& CGCG 241-068  	& 15.2   \\
 J231512.62-011458.3	& cE2 	& 23 15 12.62 		& -- 01 14 58.3 &  109.3 		&  0.83 		& 104.0  		& 504  			& III Zw 069		&  18.6   \\

\hline
\end{tabular}\label{Tab:basics}
\end{table*}

\section{Data}

Our initial search for cE candidates was taken from SDSS DR7. The main $\it{Legacy}$ survey covers $\sim$8500 square degrees of sky, with spectroscopy of various complete samples of galaxies \citep{Abazajianetal09}, Our initial set of cE candidates were selected using the SQL query language on the existing DR7 $\it{Galaxy}$ sub-sample.

The search criteria were that the galaxies have 
 half-light radii (derived from model fits, and seeing corrected by the SDSS pipeline) of
 less than 700 pc, and possess no significant emission lines  (SDSS eclass $<$ 0).
 The latter is typically chosen to select early-type galaxies (e.g. \citet{Bernardietal05}. We also require that the galaxies  have a redshift $<$ 0.025. Greater than this distance, we are unable to determine the effective radius given the mean seeing of the SDSS survey (at this distance 1$\arcsec \sim$500 pc.)
 We also only accept candidates that have distances from redshifts, as it is  known that without redshifts it is  possible to confuse an M32-type galaxy with a normal elliptical at greater distances \citep{ZieglerBender98}. Using new spectroscopic data, \cite{DrinkwaterGregg98} searched for cEs in Fornax that were listed in previous catalogues as M32-like, but all were found to be other types of galaxy. However, the decision to rely on redshifts does mean that we may miss very compact galaxies that the SDSS pipeline may not classify as galaxies, and not target for spectroscopy.

During this search for candidate compact elliptical galaxies (Huxor et al. in prep.), we have found two examples (hereafter cE1 and cE2) of compact galaxies with clearly visible tidal streams, which were identified visually during follow-up of the candidates. Both are described in detail below.

\subsection{SDSS J110404.40+451618.9 (cE1)}

\begin{figure} 
\begin{center}
 \includegraphics[angle=0,width=70mm]{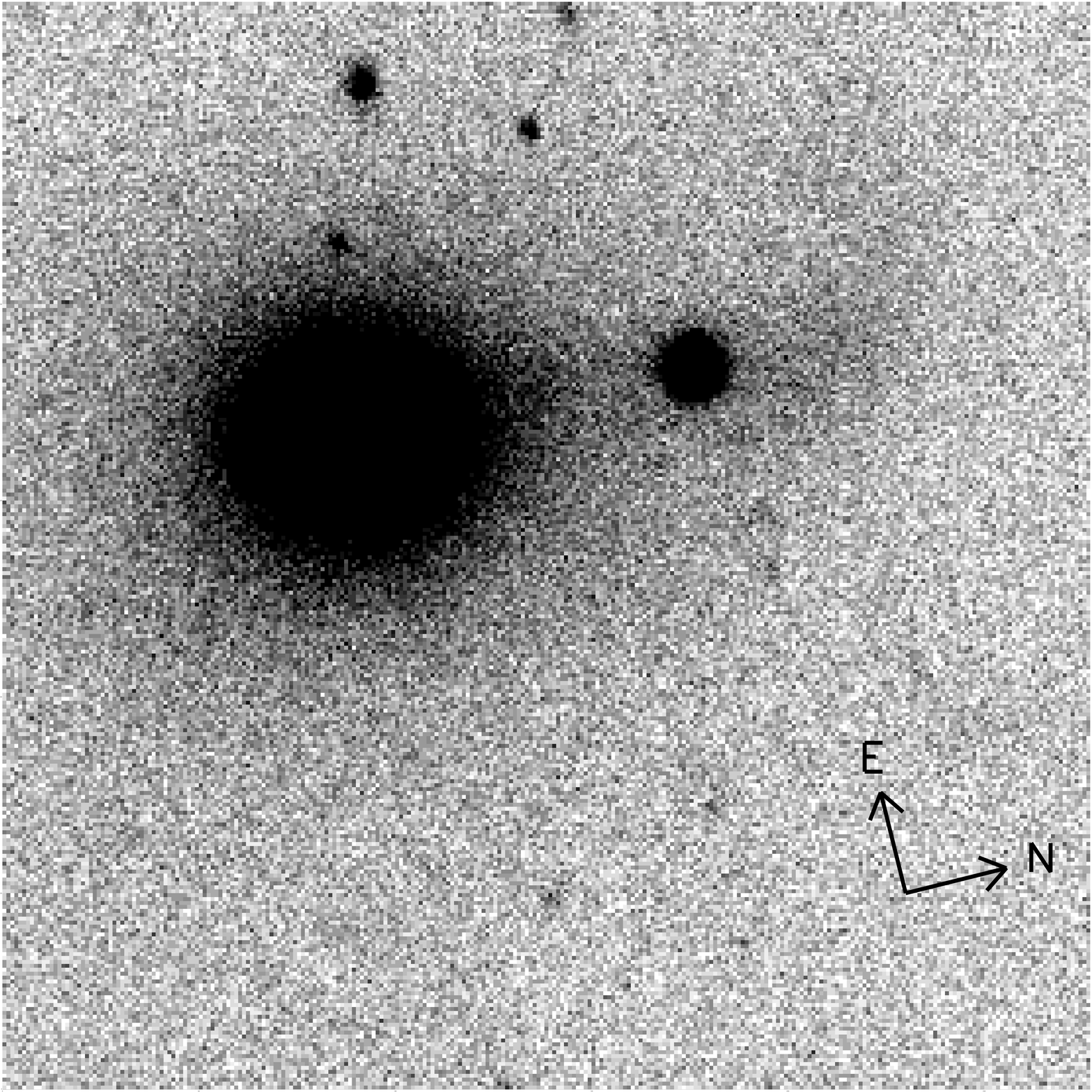}
 \vspace{1pt}
 \caption{SDSS r-band image of cE1 and its host  (CGCG 241-068), scaled to enhance the tidal tails. Image size is 50 kpc  $\times$ 50 kpc. The cE is not centred in the image, as it lies on the edge of the CCD.\label{Fi:cE1image}}
 \end{center}
\end{figure}

\begin{figure} 
\begin{center}
 \includegraphics[angle=0,width=70mm]{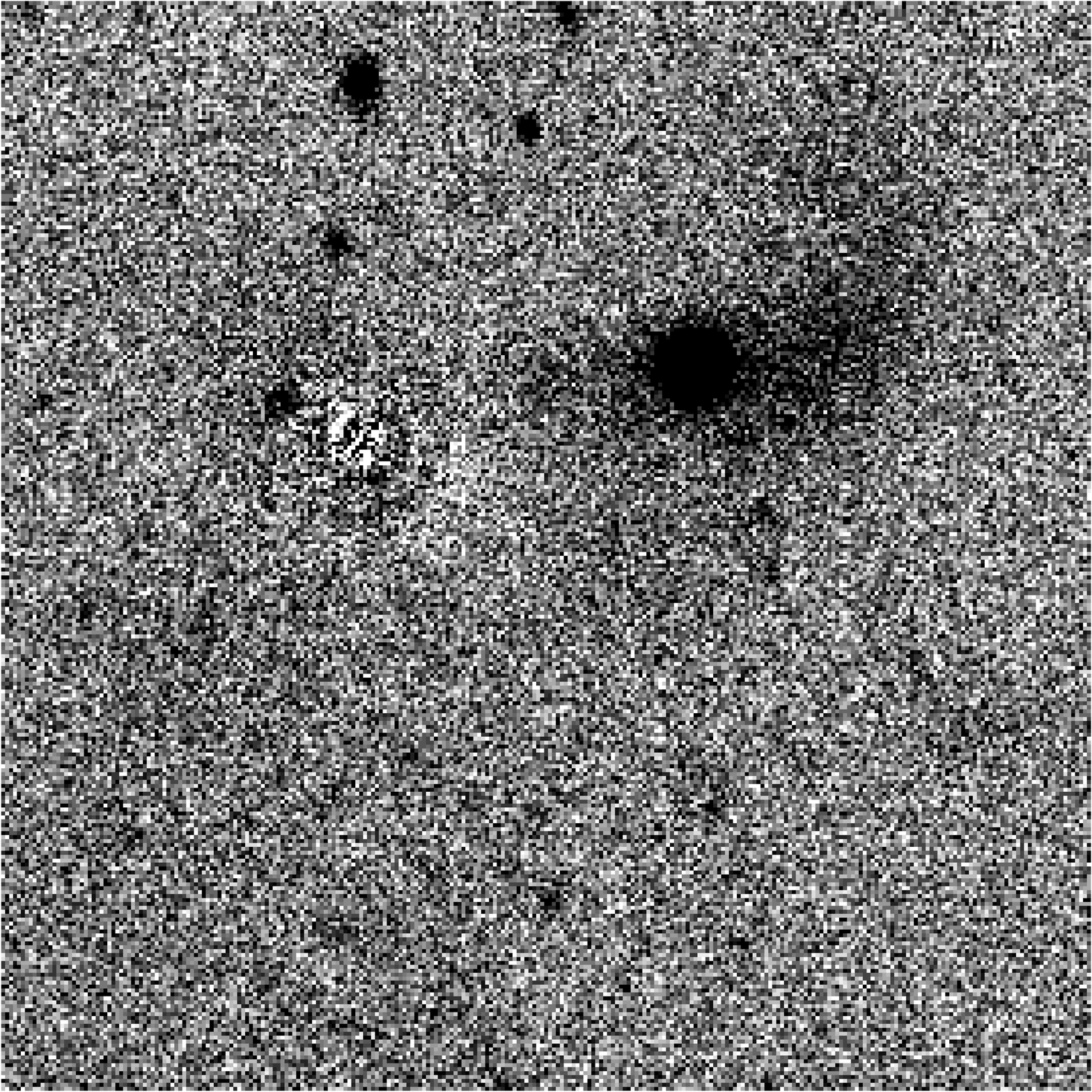}
 \vspace{1pt}
 \caption{As Figure \ref{Fi:cE1image}, but with host removed with GALFIT to show tail more clearly.\label{Fi:cE1image_nohost}}
\end{center} 
\end{figure}

Although close to its neighbouring galaxy (Figure \ref{Fi:cE1image}), a tail can clearly be seen streaming in the direction away from cE1, and a further component of the tail facing the neighbour is visible when this galaxy is subtracted with GALFIT (Figure \ref{Fi:cE1image_nohost})  The redshift of cE1 is given by SDSS as 0.0215 $\pm$ 0.0002. The neighbouring galaxy (CGCG 241-068) does not have a redshift from SDSS, but NED, using UZC data \citep{Falcoetal99}, gives a redshift for the host of 0.0220, which confirms that cE is associated with CGCG 241-068.  As some component of the cE velocity will be due to its orbital motion around the host, we take the redshift of the host to determine a luminosity distance of 95.2 Mpc. to the system\footnote{Throughout this paper we adopt a WMAP 5 year cosmology, H$_{0}$= 70.5 , $\Omega_{matter}$=0.27 
and $\Omega_{\Lambda}$=0.73, \citep{Hinshawetal09}, and use the PYTHON code version of the \citet{Wrightetal06} cosmology calculator.}, and an angular diameter distance of 91.1 Mpc. The former is used to derive absolute magnitudes, and the latter to derive effective radii, etc.

At this distance, the cE has an extinction corrected absolute magnitude, using the SDSS catalogue data, in the r-band (M$_{r0}$) of -18.57 mags, a (g-r)$_{0}$ colour of 0.81 mags, and lies at a projected distance of 15.2 kpc from the host galaxy. The host is $\sim$2.6 mags more luminous in the r band than the cE.

\subsection{SDSS J231512.62-011458.3 (cE2)}

\begin{figure}
\begin{center}
 \includegraphics[angle=0,width=70mm]{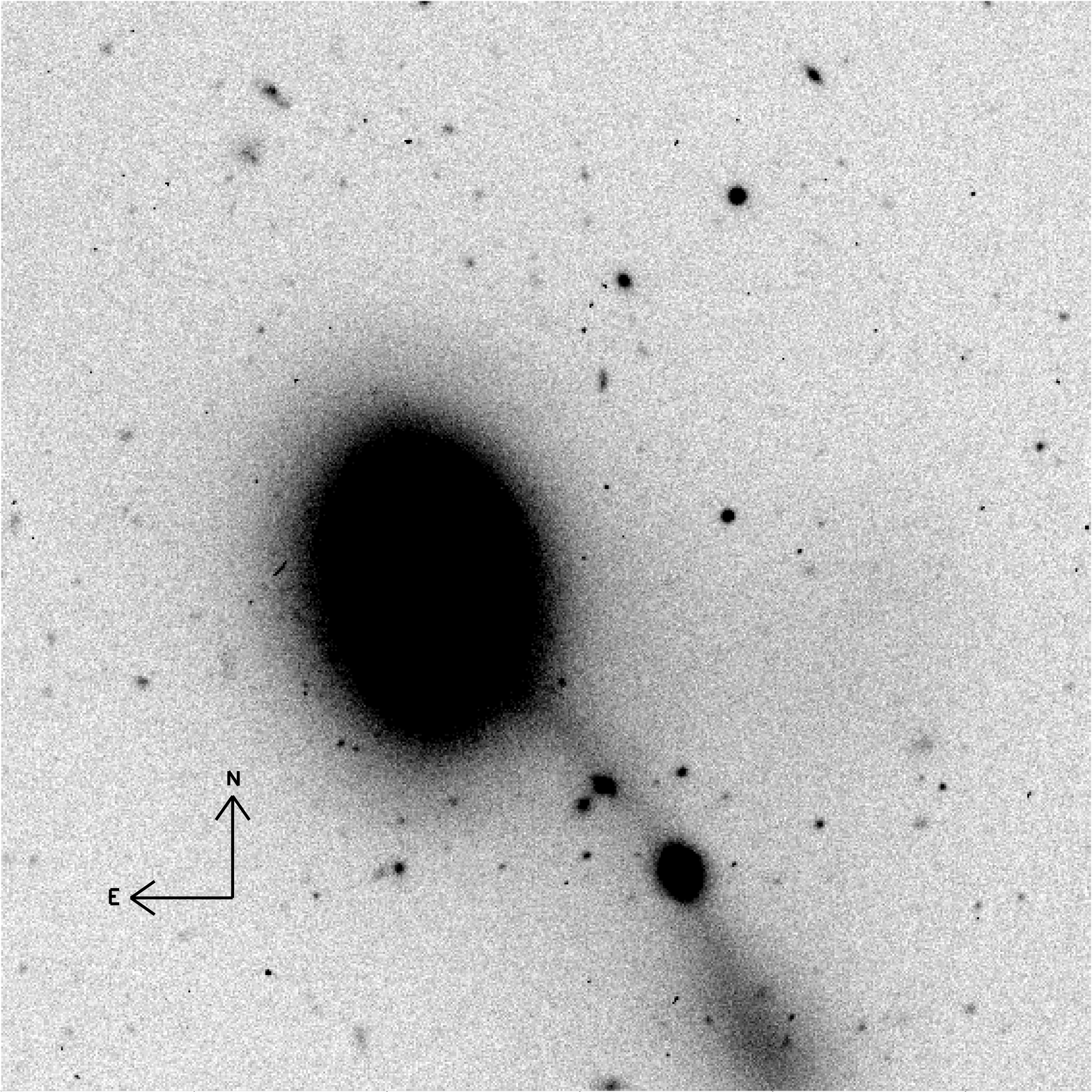} 
 \vspace{1pt}
 \caption{CFHT/Megacam r-band image of cE2  and its host galaxy (III Zw 069) scaled to enhance the tidal tails. As in Figure \ref{Fi:cE1image}, image size is 50 $\times$ 50 kpc, and it too lies on the edge of the CCD.\label{Fi:cE2image}}
 \end{center}
\end{figure}

The second cE (cE2) is a particularly fine example of stripping in  progress. In addition to the discovery SDSS image, this cE is found in archival CFHT imaging, and we have also obtained g-band imaging with WHT/ACAM. SDSS gives an M$_{r0}$ of -18.1 and (g-r)$_{0}$ colour of 0.83, and a redshift for the host of the cE as 0.0252. Using this redshift  NED gives a luminosity distance of 109.3 Mpc, and an angular size distance of 104.0 Mpc. 
With these values, the dwarf lies at a projected distance of 18.6 kpc from its host (III Zw 069) which is $\sim$3 mags more luminous than the cE. Tails can be seen from either side of the dwarf, with the southern tail  being very prominent (Figure \ref{Fi:cE2image}), and extending beyond the chip edge. A fainter diffuse structure can also be seen extending to the west of the host galaxy. 

This galaxy has considerable CFHT/Megacam imaging,  over two adjacent pointings which have a small overlap region. One CFHT pointing (Figure \ref{Fi:cE2image})  contains the cE, most of the  large tail to its south, and the host galaxy The second CFHT pointing, to the south of the first,  contains the cE and the full extent of the southern tail, but not the host galaxy (see Figure \ref{Fi:tidalRegions}).  These are much deeper than the SDSS imaging (r-band exposures of 480 seconds on the CFHT, compared to the 54 seconds on the smaller SDSS telescope).

The basic properties of the cEs, from the SDSS data, are shown in Table \ref{Tab:basics}.

\section{Analysis and Results}
\subsection{Structural Parameters}

GALFIT \citep{Pengetal02} was used to obtain revised photometric and structural parameters of the cEs (e.g. model magnitudes, effective radii and  S\'ersic parameter ``n"). These values were used in preference to the parameters provided by the SDSS pipeline. This is because the pipeline does not do a free fit on the S\'ersic parameter, but a best fit to only a de Vaucouleurs (n=4) or exponential (n=1) profile, although  S\'ersic ``n" is known to vary (e.g. \citealt{Caonetal93}). We also use GALFIT model fits to determine revised total model magnitudes of the host galaxies, to be consistent with our modeling of the cEs. The revised model fits also provide additional information on the nature of the hosts. Photometric calibration was achieved by PSF fitting to a neighbouring star(s), for which SDSS photometry is available, to obtain the appropriate zeropoint.
Where possible we use the r-band data. The values derived from the best fit models are given in Table \ref{Tab:derived}. These are also converted to B band (to allow for comparison with the literature), using the transformations in \citet{Jesteretal05}. Corrections are made for foreground reddening using the \citet{Schlegeletal98} dust maps, and k-corrections are performed using the IDL code of \citet{Chilingarianetal10}.

Errors were estimated by generating 100 realisations, based on the original images,  modified with each pixel having poisson noise on the source counts, and on random values based on the sky sigma. These were then also fit with GALFIT, and the derived  values returned used to determine the standard deviation on each parameter. In the case of cE2, we also had additional imaging data with which to estimate the likely errors. The second pointing gave an effective radius of 266 pc and a S\'ersic n of 4.93, compared to 263 pc  and a S\'ersic n of 4.36 for the first pointing. These suggest that the errors derived from the bootstrap method are underestimates, especially for S\'ersic n.  This parameter is known to be very sensitive to small changes, due to the extended wings of the profile close to the sky level, and to any possible central excess of light.

For cE1, due to the proximity of the host, both it and the cE were fit together, as the light from the outer regions was affecting the fit to the cE. With cE2, GALFIT found fitting the cE along with the host a challenge, almost certainly due to the strength and morphology of the tails.  Hence the fit was limited to a region close to the cE, to exclude much of the tails.



\begin{figure} 
 \includegraphics[angle=0,width=85mm]{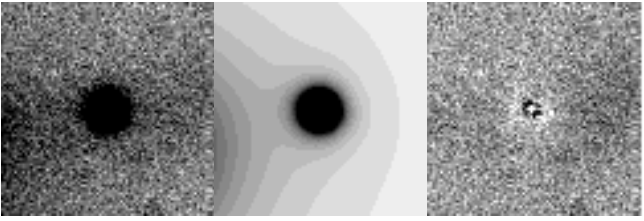}
 \vspace{1pt}
 \caption{GALFIT fit to cE1.  The original image is in the right panel, the  model in the centre, 
and the residual in the left panel. Each image is 14.0 x 14.0 kpc. The model and residual also includes that for the host galaxy. 
Orientation is the same as Figure \ref{Fi:cE1image}.}\label{Fi:ce1_galfit_figure}
\end{figure}

\begin{figure} 
 \includegraphics[angle=0,width=85mm]{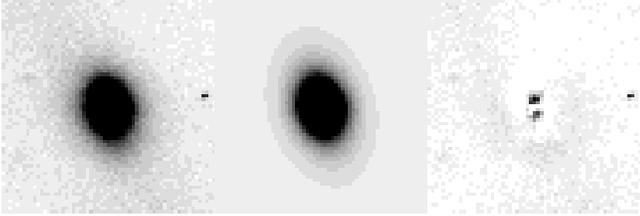}
 \vspace{1pt}
 \caption{GALFIT to cE2, similar to Figure \ref{Fi:ce1_galfit_figure}, except that the image size is 5.75 x 5.75 kpc. 
Orientation is the same as Figure \ref{Fi:cE2image}.}\label{Fi:ce2_galfit_figure}
\end{figure}

\begin{figure} 
\begin{center}
 \includegraphics[angle=0,width=85mm]{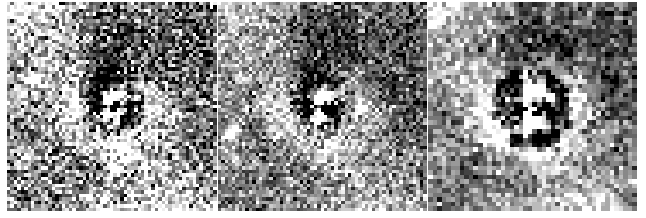}
 \end{center}
 \vspace{1pt}
 \caption{Residual images from GALFIT  fits to cE2,  showing a consistent structure. The images are all g-band to allow comparison (the only good ACAM data is g-band), and comprise (from left
 to right) Megacam at lowest part of field-of-view, Megacam from uppermost part of the field-of-view, and WHT/ACAM. North is up and east is left. The central excess  and the ring-like feature with a break in the north is also visible in all the residuals. Orientation is the same as Figure \ref{Fi:cE2image}.}\label{Fi:ce2galfitcompare}
\end{figure}

The residual for cE2 seems to possess structure, but within R$_{e}$, the residual contains $<$ 0.6 \% of the flux in the original image. 
 Figure \ref{Fi:ce2galfitcompare}  shows that the structure found in the residuals is robust. They persist in both of the  two Megacam pointings, in which the cE sits on either extreme edge of the field-of-view. We do not believe that this is an artifact of the PSF, as the Megacam PSF is strongly anisotropic \citep{Hoekstraetal06}, and is not comparable for the two locations in the left and central panels of Figure \ref{Fi:ce2galfitcompare}. We also have g-band ACAM imaging, and can thus compare this with g-band data for the Megacam. Again the same structure is apparent. There is a feature that lies roughly north-south, which maybe a bar. There is also  an excess of light in a broken torus ring around the cE, with a radius of $\sim$1 kpc.

The results  for the GALFIT  fits are given in Table \ref{Tab:derived} and illustrated in Figure \ref{Fi:plots}. Both cE1 and cE2 have effective radii and mean surface brightnesses that identify them as compact elliptical galaxies. Both have values of S\'ersic n of $\sim$4, i.e. comparable to a de Vaucouleurs profile, and as expected for either elliptical galaxies and classical bulges.

\begin{table*} Table2
 \centering
  \caption[Derived properties of cEs. ]{
 Derived properties of the cEs, derived from model fits with GALFIT, and corrected for extinction and k-correction using the code of \citet{Chilingarianetal10}, and converted to the B-band. Errors on all magnitudes are $\pm0.01$.
  }
  \begin{tabular}{@{}lllllll@{}}
 
     ID   & m$_{r}$	& M$_{r0}$	& M$_{B0}$ 			 &$R_{e}$		& S\'ersic $n$  			& SB$_{\mu_B}$   \\
            &  mags  	& mags		& mags	    			&     pc    		&				& mags arcsec$^{-2}$   \\

 \hline 
  cE1  & 16.28			& -18.65			&  --17.31    	& 388$\pm9$	& 3.58$\pm0.14$	&  19.39 	\\
  cE2  & 17.18			& -18.13			&  --16.88   	& 263$\pm1$	& 4.36$\pm0.02$	& 19.16 	\\

\hline
\end{tabular}\label{Tab:derived}
\end{table*}

\subsection{Colour maps}

\begin{figure} 
\begin{center}
 \includegraphics[angle=0,width=70mm]{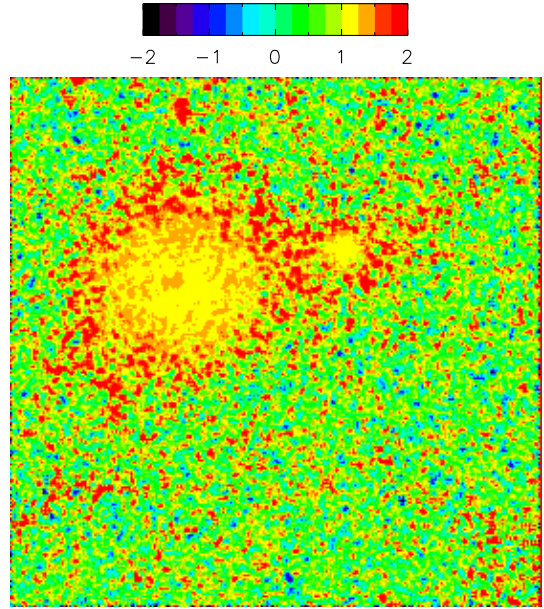}
 \vspace{1pt}
 \caption{Colour map in (g-i)$_{0}$ for cE1, using extinction values for g and i-band obtained from SDSS. The original image was smoothed with a boxcar average of 3 $\times$ 3 pixels. Due to the faintness of the tidal feature, it is barely visible in the colour map. 
Note that the host and cE have similar colours.\label{Fi:cE1colourmap}}
 \end{center}
\end{figure}

\begin{figure} 
\begin{center}
 \includegraphics[angle=0,width=70mm]{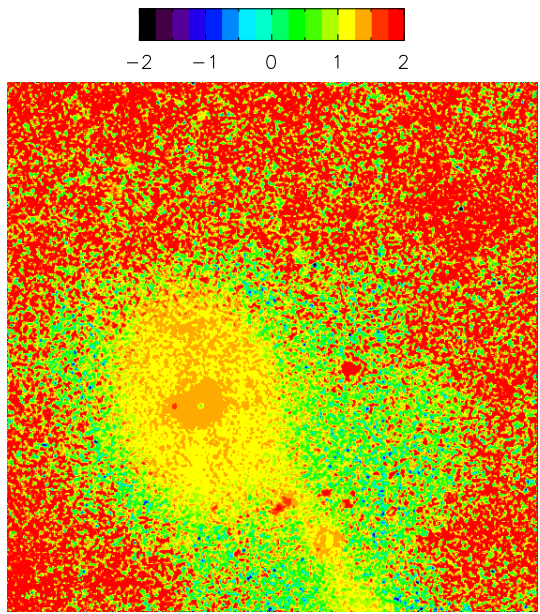}
 \vspace{1pt}
 \caption{Colour map in  (g-i)$_{0}$ for cE2. As in Figure \ref{Fi:cE1colourmap}, the original image was smoothed with a boxcar average of 3 $\times$ 3 pixels. The centre of the host galaxy is saturated in the i-band leading to the central feature. A redder region around this central feature probably represents the bulge of the host, having a radius of $\sim$1.5 kpc. The blue region surrounding the two galaxies can be seen in the g-band images, and is suggestive of a stripped disk either
 from the host, or from the disruption of other satellites. The sky is redder than in Figure \ref{Fi:cE1colourmap} as the i-band image has an exposure time that is about twice that of the g-band for the Megacam data, whereas the SDSS imaging of cE1 has equal exposures in all filters. \label{Fi:cE2colourmap}}
 \end{center}
\end{figure}

Colour maps of both cEs, and their hosts were generated by taking sky-subtracted images, and converting the flux into surface brightness employing the zeropoint calibration obtained previously. These are shown in Figures \ref{Fi:cE1colourmap} and \ref{Fi:cE2colourmap}. 

In the case of cE1, the signal from the tidal features is too faint, and do not show in the colour map. For cE2, the tail to the south is bluer than the cE itself. The diffuse region seen in Figure \ref{Fi:cE2image} is also clearly visible in the colour map, and is much bluer than the bulk of cE2 or its host suggesting either a younger or more metal-poor stellar population. It is notable that there is the suggestion of  redder colour in the outer region of cE2, at $\sim$1 kpc from its centre, which may be associated with the excess of flux found in the GALFIT residuals noted above.

\subsection{Surface Photometry of Tidal Features}

\begin{figure} 
\begin{center}$
\begin{array}{cc}
\includegraphics[angle=0,width=40mm]{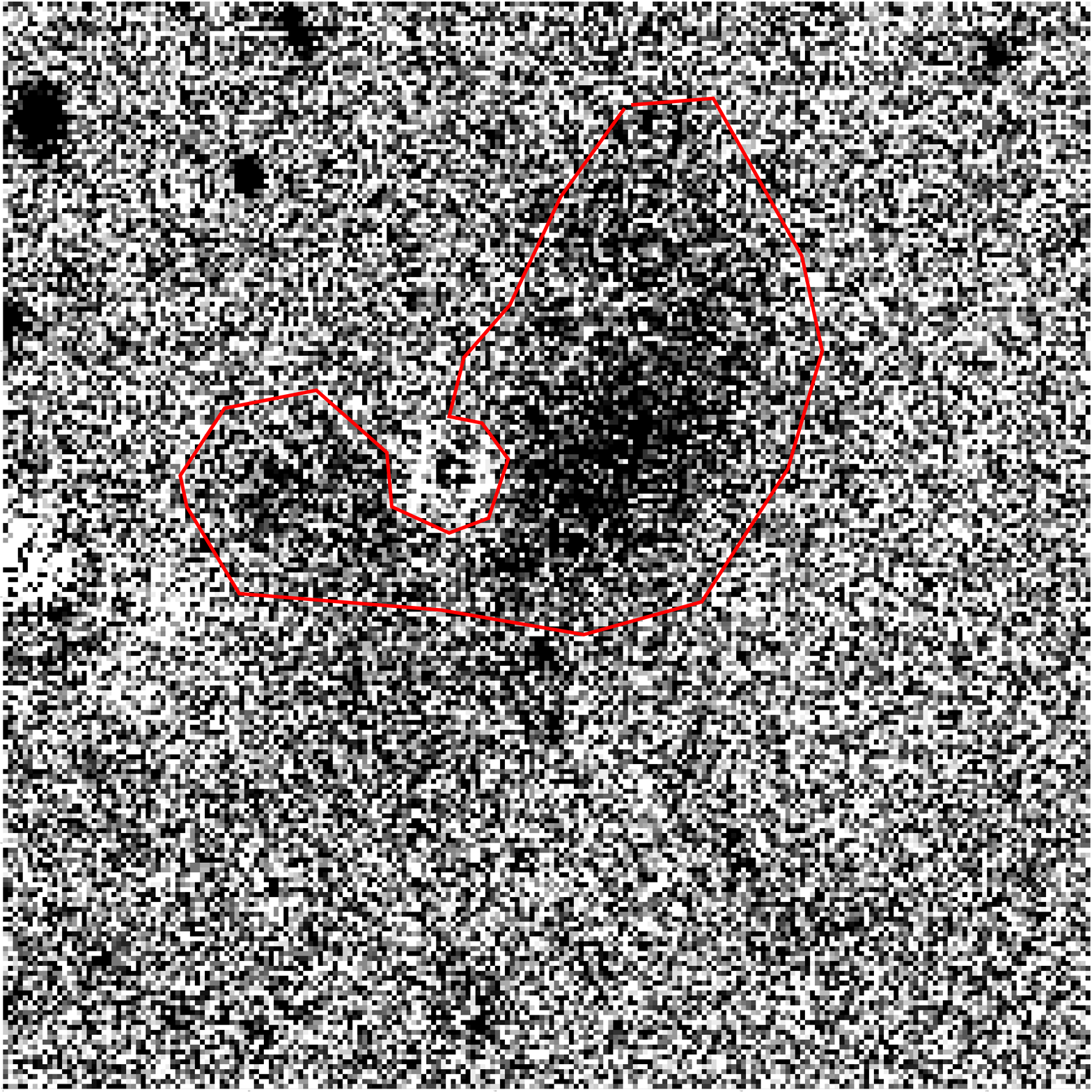} &
\includegraphics[angle=0,width=40mm]{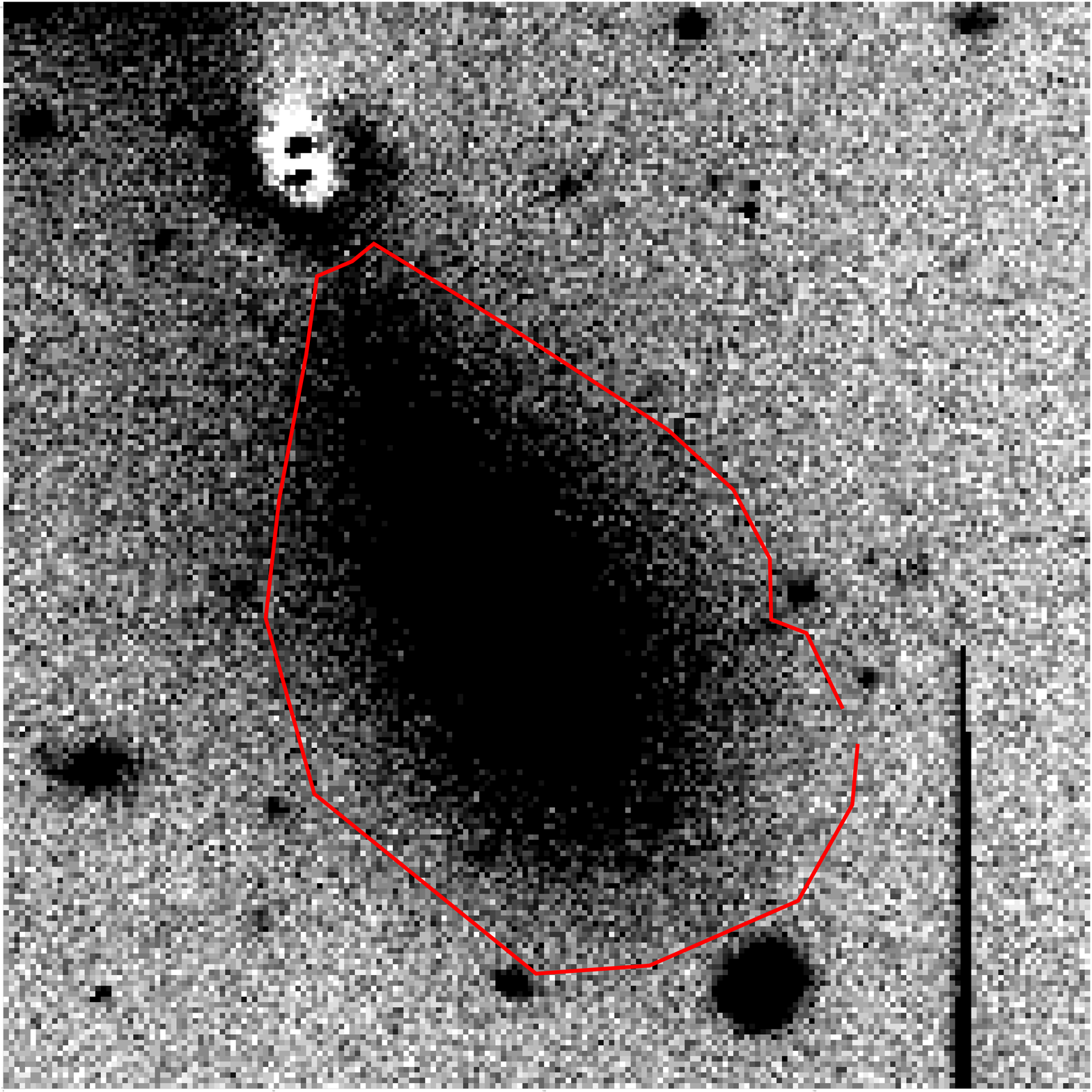}
\end{array}$
\end{center}
\caption{The regions used (in red) for surface photometry (see Table \ref{Tab:tails}) of the main tidal features in cE1 (left) and cE2 (right), plotted over the residual images. 
The areas are 1153 arcsec$^{2}$ (cE1) and 1534 arcsec$^{2}$ (cE2). 
The images are 38$\times$38 kpc in size for cE1 and 19$\times$19 kpc for cE2, and have been scaled to highlight the tails.\label{Fi:tidalRegions}}
\end{figure}

Regions of the tidal structures were selected (see Figure \ref{Fi:tidalRegions}) to determine the magnitude of the tidal features. The regions were defined using the residual images from the GALFIT fits, which had also been smoothed over 3 pixels with a Gaussian, to increase the S/N of the diffuse features. The residual images (from the GALFIT fits to the original unsmoothed data) were used for photometry, to minimise the effect of the background halo light from the host galaxies. This was particularly important for cE1, which lies within the halo of CGCG 241-068.  The results are given in Table \ref{Tab:tails}. The magnitudes of the tails are probably lower limits, as in cE2 only  the tail to one side of the dwarf could be measured, as the side facing  the host galaxy is lost in background light. Moreover, there is also likely stellar material in the tails at very low surface brightness limits.

The errors on the magnitudes for the tails were derived, following \citet{Sasakietal07}, using the relation:
\begin{equation}
\sigma^2(total$\_$counts) = total$\_$counts + N\sigma^2_{background} 
\end{equation}

where N is the total number of pixels in the selected tail region. For cE1, the errors on the magnitude of the tail is $\pm0.01$ mags. For cE2, the formal error is better, at $\pm0.001$. However, the value of the magnitude is affected by the area selected to represent the background sky. There is considerable  diffuse light around cE2 and its host, best seen in Figure \ref{Fi:cE2colourmap}. We have selected a region beyond this diffuse light to obtain the sky. If this extra light is not from cE2, then a proportion of the flux attributed to the cE may be from another source.

In both cEs a large proportion of the stellar mass is found in the tails: more than a third in the case of cE1, and about two-thirds in cE2. But as noted above, these are lower limits.
 
\begin{table*} 
 \centering
  \caption[Properties of tidal features]{
Properties of tidal features, as defined by the red regions in Figure \ref{Fi:tidalRegions}, compared to GALFIT fit derived value for cEs. See text for discussion of errors. }
  \begin{tabular}{@{}lllllll@{}}
     ID		&    m$^{tail}_{r}$ 	& M$^{cE}_{r0}$ 	&  M$^{tail}_{r0}$  	& M$^{cE+tail}_{r0}$		& flux ratio  	& SB$_{\mu_r0}$  \\
          	&     mags       		& mags				& mags 					&	mags		        	&  tail/cE 		& mags arcsec$^{-2}$    \\

 \hline 
  cE1 		&  17.34				& -18.65  			&  -17.57				& -18.99					& 0.37 				&   25.0		\\
  cE2 		&  17.61				& -18.13  			&  -17.69 				& -18.68					& 0.67  			&   25.6		\\

\hline
\end{tabular}\label{Tab:tails}
\end{table*}

\subsection{Spectra}
We use the SDSS spectra for each of these galaxies, using the approach described in \citet{Priceetal09} to determine ages and key indicators and their errors as described in  \citet{Priceetal09,Priceetal10}, but we summarize briefly here. We use the models of \citet{Schiavon07} and the EZ-Ages code devised by \citet{GravesSchiavon08} to measure the age, metallicity and key abundance indicator [Mg/Fe] of our compact galaxies. 
These are the values of the parameters for a SSP (single stellar population) which most closely reproduce the data.  Velocity dispersions are from SDSS.

It is notable that both cEs have young to intermediate SSP equivalent ages.

\begin{table} 
 \centering
  \caption[Spectroscopic Properties of cEs]{
Spectroscopic Properties of cEs
  }
  \begin{tabular}{@{}lllllllll@{}}
     ID 	& Age (Gyr)	 	&  [Fe/H]			& [Mg/Fe] 			 & $\sigma$ (kms$^{-1}$)  	 \\
 \hline 
  cE1	 	& 5.33$\pm1.52$	& 0.14$\pm0.18$	& 0.12$\pm0.16$ 	&  89$\pm6$   				 \\
  cE2		& 5.40$\pm1.64$	& -0.01$\pm0.24$	& 0.07$\pm0.22$	&  92$\pm10$ 				 \\
\hline 
\end{tabular}\label{Tab:spectral}
\end{table}


\begin{figure*} 
\begin{center}$
\begin{array}{cc}
\includegraphics[angle=90,width=87mm]{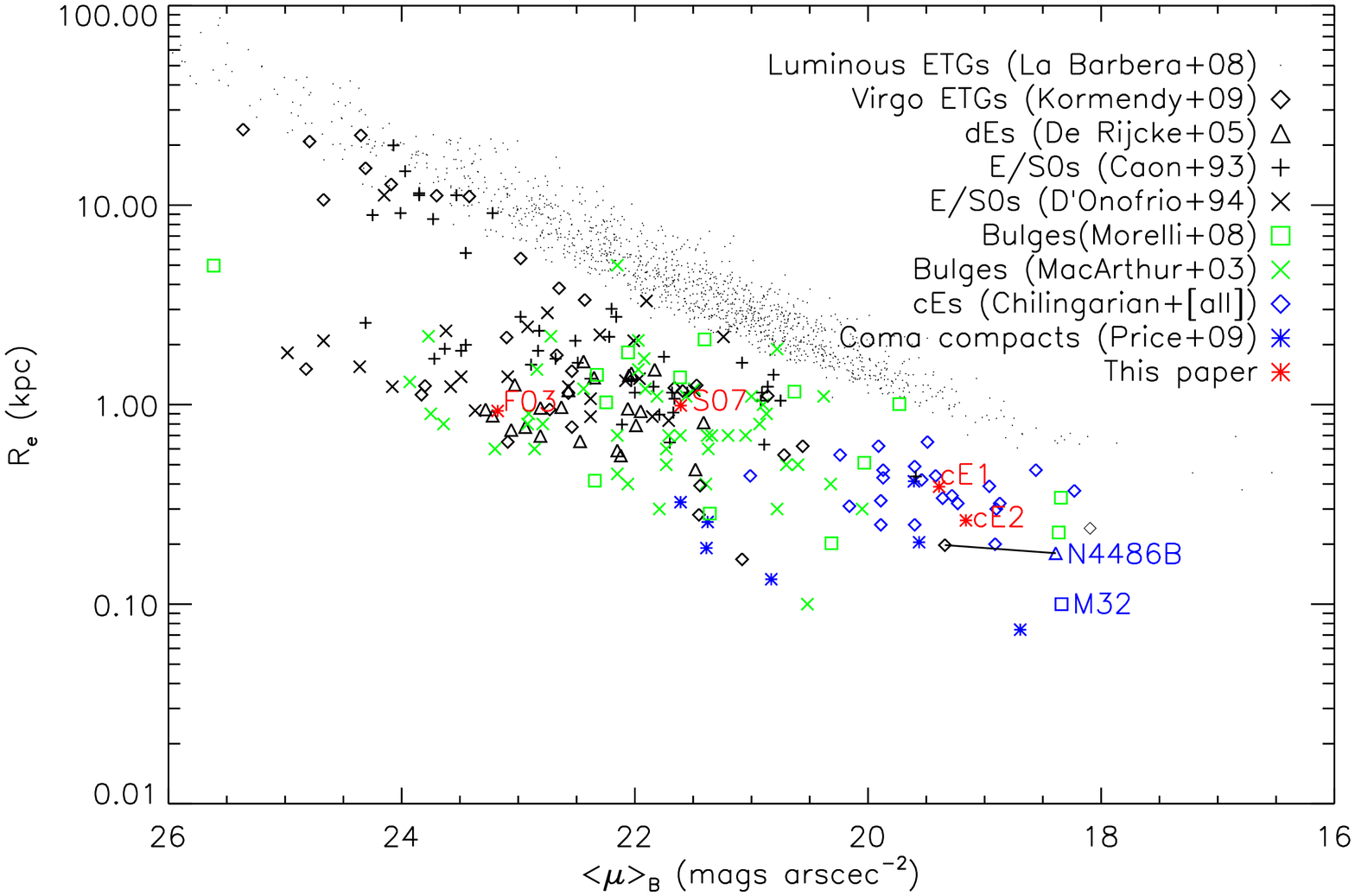} &
\includegraphics[angle=90,width=87mm]{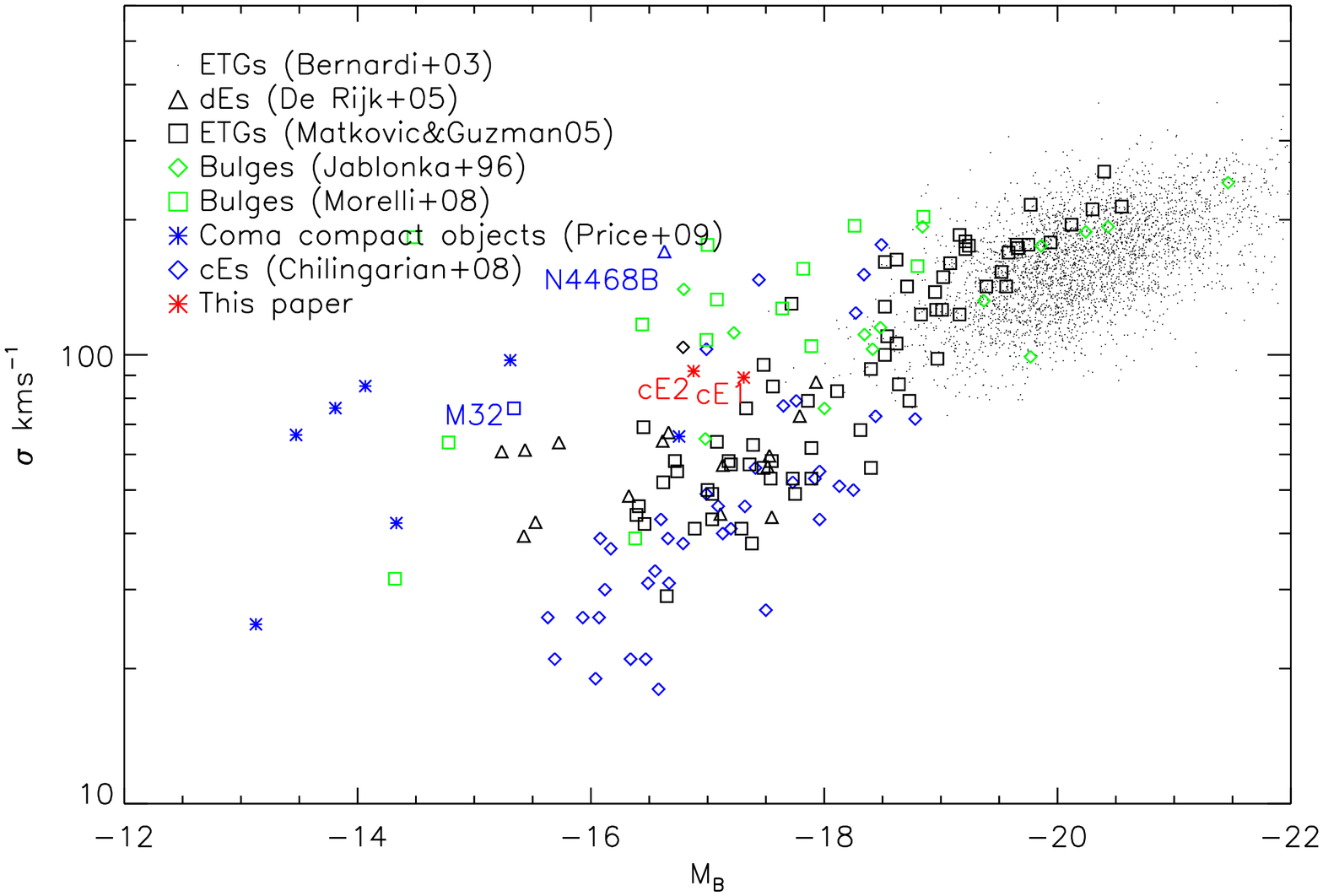}
\end{array}$
\end{center}
\caption{Relation of the new cEs to other hot stellar systems. The new cEs (cE1 and cE2) are labelled.The dwarfs of \citet{Forbesetal03} and \citet{Sasakietal07} are also shown as red and labelled F03 and S07. Published cEs are blue, ETGs are black, and bulges green.  Different datasets with these are shown by different symbols, and all are converted to our chosen cosmology (5 year WMAP).  
Left: Plot of stellar systems, including ETGs, bulges and known cEs in surface brightness vs effective radius. Also shown are the galaxies suffering stripping
 from \citet{Forbesetal03} and \citet{Sasakietal07}. For the data of \citet{LaBarberaetal08} the equations in \citet{Bernardietal03} are used to derive $\mu_{g}$ and $\mu_{B}$ from the relations in \citet{Jesteretal05}. Colours, allowing the determination of B-band magnitudes for the \citet{Kormendyetal09} data were obtained from the GOLDMine database (goldmine.mib.infn.it), where available. We also exclude the galaxies classified as S0 in the \citet{Kormendyetal09} sample. Values for M32 and NGC 4486B from \citet{Chilingarianetal07} are labelled, with a black line connecting the data points for NGC 4486B from \citet{Chilingarianetal07} and \citet{Kormendyetal09}.
		Right: The Faber-Jackson relation. We include data from \citet{Bernardietal03} for a redshift $< 0.1$, with colour conversion using \citet{Jesteretal05}, \citet{MatkovicGuzman05} (where S/N $>$ 15),
 \citet{Chilingarianetal08} (for early type galaxies in their sample only), \citet{Jablonkaetal96} for bulges, \citet{Morellietal08}, where their R-band magnitudes are converted to B using a B-R = 1.45 as found in \citet{BalcellsPeletier94}.} \label{Fi:plots}
\end{figure*}

\subsection{Host Galaxies and Environments}

\subsubsection{cE1}

The host of cE1(CGCG 241-068) was best fit by GALFIT using three components, all of which had a S\'ersic n of $\sim$4, and a range of effective radii. It was this fit that was used to subtract the host for the GALFIT fit to the cE. However, a single component with  an n of 5.47 gave a good fit, consistent with the host being a classical elliptical galaxy. 

cE1 is in the \citet{FocardiKelm02} catalogue of compact groups (group number 127). In this catalogue, which only lists four members in the group, the host is the only absorption line galaxy the others also having emission features.  
It is also in the group U349 from \citet{Ramellaetal02}, which contains an additional two members. The majority of the galaxies in the  \citet{Ramellaetal02} group remain late types, with the cE host being the only clear massive elliptical galaxy.  It is notable that $\sim$2 arcmin ( about  55 kpc in projected distance) north of cE1 there is a more massive galaxy (PGC 033427), whose image is also showing strong signs of being stripped by CGCG 241-068 (the host of cE1). That is, this cE host is not only currently stripping two other galaxies, but has most likely had some previous history of merging activity to obtain its current  early-type morphology. In the standard hierarchical model of galaxy formation massive ellipticals are the result of mergers.

\subsubsection{cE2}


Although Megacam imaging is used to study the dwarf galaxy, the r and i-band images are so deep that the host galaxy (II Zw 097)  is saturated in its central regions. Therefore we employ the SDSS data to study it. The GALFIT fit to the SDSS r-band image gives a minimal residual with a two S\'ersic and central PSF fit to a nuclear component. The inner S\'ersic component has an effective radius of $\sim$1 kpc., and the B/T ratio is $\sim$0.5. The outer component has a low S\'ersic n of 0.45 and a effective radius of 5.5 kpc. This is consistent with a S0/a type galaxy \citep{Oohamaetal09}.

It is hard to say much about the environment of cE2 as it lies  only 1 $\arcmin$ ($\sim$30 kpc)  from the edge of the SDSS DR7 survey area. The SDSS image does not go further south than the cE, but to the north of its position the SDSS database gives a further four galaxies at this redshift range, all of which are late type. It also  gives a galaxy, visible in the Megacam imaging (but not the SDSS images) as a disk galaxy, at the same redshift as cE2, and at a projected distance of $\sim$300 kpc.  All of these for which spectra are available show evidence of strong emission.  cE2 is also near a \citet{Ramellaetal02} group (U834) both on the sky and in redshift. This group has only three listed members, the closest of which is $\sim$38 arcmin from III Zw 097, or 1.1 Mpc. This is sufficiently far that the cE2 system is described as isolated in \citet{Pradaetal03}. The SDSS spectroscopic data further shows that the early type galaxies nearby on the sky are members of a background galaxy cluster at a redshift of $\sim$0.1, and so not associated with the group.
  
The large spiral fraction of these groups is not unexpected, as they are relatively small possessing only a handful of galaxies. However, it is notable that the two cEs each  are close to  an early-type host galaxy.

 \section{Discussion}

\subsection{Pre-processing}

One notable feature of both cE galaxies is the environments in which they are found -- relatively small groups. At first sight  this is surprising, as stripping was expected to be found in galaxy clusters for a number of reasons. Firstly, previous searches in small group environments, such as the Leo Group  \citep{ZieglerBender98}, had proved unsuccessful, with candidates being shown not to be cEs. And secondly, many had argued that the dense environments of galaxy clusters were the best location to find cEs (e.g. \citealt{Mieskeetal05}). Hence many recent searches have focussed on galaxy clusters (e.g. \citet{Chilingarianetal09}). The new cEs presented here, however, show that cEs can form in small groups. However, this should not be surprising as the prototype cE, M32, is in our own Local Group, which only comprises a few luminous late-type galaxies (MW, M31 and M33).

An outstanding question is whether cE1 and cE2 are more representative of cEs than those that have been found previously in galaxy clusters, or whether they are unusual (in being located in small groups and exhibiting tidal features) but simply easier to detect. We may only be seeing a result of a number of effects. Firstly, there are considerably  more galaxies in groups than in clusters \citep{Ekeetal05}, so one would expect to see instances in groups given a search of a wide-field survey such as SDSS DR7. Secondly, the crowded nature of galaxy  clusters may reduce the time that a stream  is visible as associated with its satellite galaxy.  Simulations \citep{Rudicketal09} show that early in a galaxy cluster's life, when the majority of the galaxies are found in smaller groups that have yet to merge into the final single cluster, stream production dominates the formation of the  intracluster light (ICL). Later, the heating of the stellar material in galaxies in the cluster, and the complex tidal field, mean that streams are rapidly dissolved.  Finally there may be much background light from the cluster members and its ICL, making the identification of streams, if present, more difficult. The less harsh  smaller group environments are thus ideal to detect stripping and also to study the details of the stripping process. 

However, the presence of the two new cEs in groups could be a real effect, and there may be a preference for cEs to form in group rather than cluster environments. Although the overall tidal field may be less in a group than a cluster, the lower relative velocities of the galaxies within a group allow time for stripping to have an effect.

This scenario would be consistent with the notion of ``pre-processing", in which much more of the transformation of galaxies, than hitherto expected, occurs when they are in groups and not in any subsequent cluster environment. Preprocessing has been proposed to explain a number of recent observations. For example,  \citet{Wilmanetal09} conclude that group or subgroup environments are the main sites  of S0  formation, and that minor mergers, harassment and tidal interactions are the most likely mechanisms. Similarly, \citet{Thomasetal08} found that a simple threshing model for UCDs \citep{Phillippsetal01} within a galaxy cluster potential failed, in that it produced insufficient UCDs at large cluster-centric radii, compared to the data from the Fornax Cluster. But they suggest that this may be due to the UCDs having been formed in sub-groups  that later fell into, and became, Fornax. \citet{Kautschetal08} find that the four individual groups being assembled into the galaxy cluster SG 1120-1202 have early-type galaxy (ETG) fractions consistent with  pre-processing occurring. Indeed they argue that it plays a dominant role in establishing the ETG fraction of galaxy clusters. A similar conclusion was drawn by  \citet{BaloghMcGee10} from a study of a large sample of galaxies from SDSS.

It was noted above that the groups containing our cEs have a large spiral fraction, indicating that the environment is unevolved, even when compared to other  galaxy groups. This is consistent with the idea that the youth of the group allows  one to see the initial stages of cE formation, signatures which will be erased in a couple of Gyr or so, and hence not visible in older, more evolved, groups.

Many cEs have now been found \citep{Priceetal09,Chilingarianetal09}, but none of these exhibit strong tidal features such as seen above, although there are hints of halos and or other more diffuse features. It may be that these compact galaxies formed in groups before they assembled into a massive galaxy cluster. This might explain the old ages of the cEs found in galaxy clusters  \citep{Chilingarianetal09, Chilingarianetal07}, which may have been transformed into cEs many Gyr ago, prior to assembly in the cluster.

\subsection{Nature of the cE progenitors}

The progenitors for the two cEs were almost certainly disk galaxies, as computer simulations (e.g. \citealt{Feldmannetal08})  indicate that discs are required to get tidal tails. The simulations of \citet{HigdonWallin03} show that the teardrop morphology, similar to that seen in  cE2,  is best explained by a rotating disk in the dwarf galaxy, and occurs, in their model, at $\sim$ 1Gyr after initial impact of the dwarf on the host galaxy. 
Indeed, the arm and loops are only formed from the disk components, and these also have longer lifetimes ( $\sim$1 Gyr, up to 3 Gyr for the less strong features) than the much weaker features (typically broad fans) from elliptical-elliptical interactions. Other N-body simulations \citep{Bekkietal01b,Chilingarianetal09} further show that the stripping of a disc galaxy can specifically lead to cE galaxies. With cE1 and cE2, we have direct evidence of this process occurring.

In essence, the cEs are the bulges of the progenitor disc galaxies. They can be understood as essentially 'naked bulges', sharing most of the properties of the bulges of their progenitors, although some changes are expected from the interaction.  A focus on a bulge origin for the bulk of these galaxies is important, allowing us to interpret the location of cEs with respect to the Kormendy relation and the Faber-Jackson relation (see Figure \ref{Fi:plots}). In these plots, they sit in the regions occupied by bulges, and no longer appear to be so unusual.

\subsection{Ages}

Both cE1 and cE2 possess  SSP equivalent ages that are relatively young compared to the majority of elliptical galaxies.
It is notable that the presence of a younger stellar population makes the new cEs somewhat more akin to the the template cE (M32) than others reported in the literature.  \citet{Schiavonetal04} give a spectroscopic age for M32 of between 2.0 and 3.5 Gyr, compared to SSP ages of  $\sim$5.4 Gyr for cE1 and cE2, whereas \citet{Chilingarianetal09} find that none of their cEs exhibits a young stellar population, and the majority are old ($\geq$9 Gyr).  cE1 and cE2 are, however, not unlike the \citet{Forbesetal03} galaxy noted earlier. It has broad-band colours of  B-V=0.34, and (V-I) = 0.5, and these  blue colours  suggested to them that a population of young stars is present. Stellar population  models  gave a mean [Fe/H] of -0.7, and a luminosity-weighted age of 2 $\pm$ 1 Gyr.

In the case of M32, \citet{Coelhoetal09} analysed high S/N spectra and found both an ancient and an intermediate-age population, with the latter being more concentrated in the nuclear region. This is consistent with the stripping model in which the starburst is caused by gas inflow to the centre of the smaller galaxy. For \citet{Bekkietal01b} found that, in  their simulation of the formation of M32 by stripping, rapid gas inflow towards the central regions of the dwarf occurs, giving rise to efficient star formation. 
One might expect star formation to be triggered by the gravitational interaction.  \citet{WoodsGeller07} investigated minor galaxy interaction in SDSS DR5. They find that the minor partner in the interaction  frequently suffers triggered star formation, and that, as expected, the gas-rich galaxies experience this more. However, it is also consistent with the view of \citet{Kormendyetal09} in which a starburst results from a recent merger that formed M32. It is the clear visual evidence of tidal streams that shows the stripping model to the be correct one for our cEs.

\section{Conclusion}

We have presented two newly discovered compact elliptical (cE) galaxies, which show clear evidence for ongoing tidal stripping, and are being transformed from progenitor disk galaxies. These galaxies show that tidal stripping is one channel for the formation of compact ellipticals -- although it does not exclude the possibility of other channels being available.  Both the new cEs have a younger stellar population than most ETGs, likely triggered by the tidal interactions. 

We previously have found circumstantial evidence of stripping from the metallicity and velocity dispersions of recently discovered compact galaxies in the Coma cluster \citep{Priceetal09}. For the first time we see, in the galaxies presented here, direct evidence for this process and do not need to rely on the indirect evidence of galaxy properties lying off the scaling relations for spheroidal stellar systems

Both are also in relatively sparse environments, in small groups. It is also notable that the hosts of the cEs are the only early-type galaxy in their local environments, although no immediate explanation for this presents itself. 

The two new cEs are much like the prototype cE, M32. They share many of its properties, including a young stellar population, and being located in a low density group. In the case of cE2, its small size and environment make it more akin to M32 than the other cEs in the literature.

It is valuable to have such examples of the process of tidal stripping of discs. Not only do they allow detailed study of the process of cE formation, but also stripping is believed to be an important channel for the formation of other types of galaxy in addition to cEs, such as UCDs \citep{Goerdtetal08, Bekkietal01a}, and dEs \citep{Mooreetal98}. Follow-up studies of these new cEs will allow a better understanding of some of the details of the process, and the environments in which it occurs.

\section*{Acknowledgments}

APH would like to acknowledge the generosity of the Leverhulme Trust, during the course of whose grant this work was undertaken.

The WHT and its service programme are operated on the island of La Palma by the Isaac Newton Group in the Spanish Observatorio del Roque de los Muchachos of the Instituto de Astrof\'{i}sica de Canarias.
This work is partly based on observations obtained with MegaPrime/MegaCam, a joint project of CFHT and CEA/DAPNIA, at the Canada-France-Hawaii Telescope (CFHT) which is operated by the National Research Council (NRC) of Canada, the Institute National des Sciences de l'Univers of the Centre National de la Recherche Scientifique of France, and the University of Hawaii.
Funding for the SDSS and SDSS-II has been provided by the Alfred P. Sloan Foundation, the Participating Institutions, the National Science Foundation, the U.S. Department of Energy, the National Aeronautics and Space Administration, the Japanese Monbukagakusho, the Max Planck Society, and the Higher Education Funding Council for England. The SDSS Web Site is http://www.sdss.org/.
This research has made use of the NASA/IPAC Extragalactic Database (NED) which is operated by the Jet Propulsion Laboratory, California Institute of Technology, under contract with the National Aeronautics and Space Administration.
We also thank the anonymous referee whose comments improved the presentation of this paper.

\bsp

\label{lastpage}

\end{document}